\documentclass[amsmath,amssymb]{elsart} 
\usepackage{graphicx}
\usepackage{amssymb}
\usepackage{times}
\usepackage{dcolumn} 
\usepackage{bm} 
\usepackage{txfonts}
\usepackage{multirow}

\begin{document}

\begin{frontmatter}

\title{Light-weight Flexible Magnetic Shields For Large-Aperture Photomultiplier Tubes}

\author[uc,lbl]{P. DeVore},
\author[lbl]{D. Escontrias},
\author[uici]{T. Koblesky},
\author[lbl]{C. J. Lin},
\author[lbl]{D. W. Liu},
\author[uc,lbl]{K. B. Luk},
\author[lbl,cuhk]{J. Ngan},
\author[uici]{J. C. Peng},
\author[uici]{C. Polly},
\author[uici]{J. Roloff},
\author[uc,lbl]{H. Steiner},
\author[uc,lbl]{S. Wang},
\author[lbl,cuhk]{J. Wong},
\author[bnl]{M. Yeh}
\address[uc]{Department of Physics, University of California, Berkeley, CA
94720, U.S.A.}
\address[lbl]{Physics Division, Lawrence Berkeley National Laboratory,
Berkeley, CA 94720, U.S.A.}
\address[uici]{Department of Physics, University of Illinois at Urbana-Champaign, Urbana, IL 61801, U.S.A.}
\address[cuhk]{Department of Physics, Chinese University of Hong Kong, Hong Kong, China}
\address[bnl]{Chemistry Department, Brookhaven National Laboratory. Upton, NY 11973, U.S.A.}

\date{August 20, 2013}


\begin{abstract}
Thin flexible sheets of high-permeability FINEMET${}^{\textregistered}$ foils encased in thin plastic layers have 
been used to shield various types of 20-cm-diameter photomultiplier tubes from ambient magnetic fields.  In the presence of the Earth's magnetic field this type of shielding is shown to increase the collection efficiency of photoelectrons and can improve the uniformity of response of these photomultiplier tubes.     
\end{abstract}

\end{frontmatter}

\section{Introduction}
It is well known that the performance of most photomultiplier tubes (PMTs) is susceptible to even small local magnetic field such as that of the Earth, which is typically in the range of 0.4 to 0.5 gauss \cite{pmt_hdbk}.  The impact of this kind of weak magnetic field can be reduced by shielding the PMT, especially the region near the photocathode, with high-permeability materials such as mu-metal.   


We describe here a new type of magnetic shield that has been developed at the Lawrence Berkeley National Laboratory for use in the Daya Bay Reactor Neutrino Oscillation Experiment \cite{dayabay}. 
It utilizes FINEMET, a novel thin flexible high-permeability material, to 
form a truncated cone around a large-aperture PMT behind the photocathode.  In the following sections we describe the detailed characteristics of this shield, and present experimental results showing its impact on the performance of some selected PMTs.  In particular we will show that it reduces variations in the PMT response caused by changes in the relative orientation of the PMT and magnetic field, and that it significantly improves the electron collection efficiency of the PMT.  These improvements have a direct impact on the physics objectives of an experiment in that they result in a greater uniformity of response of the detector, and thereby yield a more reliable measurement of the energy of an event.


\section{Effect of Magnetic Field on PMT}
We have investigated the effect of the magnetic field on the gain and charge collection of
some large-aperture PMTs. These results are of particular relevance to experiments where the events of interest produce at most a few photoelectrons (PEs) in a given PMT.  The magnetic field also affects multi-photoelectron events, but in this case it is not possible to separate the effects caused by gain variation from those attributable to collection efficiency.  In this study we used Electron Tubes 9354KB \cite{et9354}, Hamamatsu R5912 \cite{hama5912}, and Photonis XP1806  \cite{xp1806} 20-cm PMTs. 

\subsection{Effect of Magnetic Field on Collected Amount of Charge}
\begin{figure}[htpb]
\centering
\includegraphics[height=3in, width=0.8\textwidth, keepaspectratio=true]{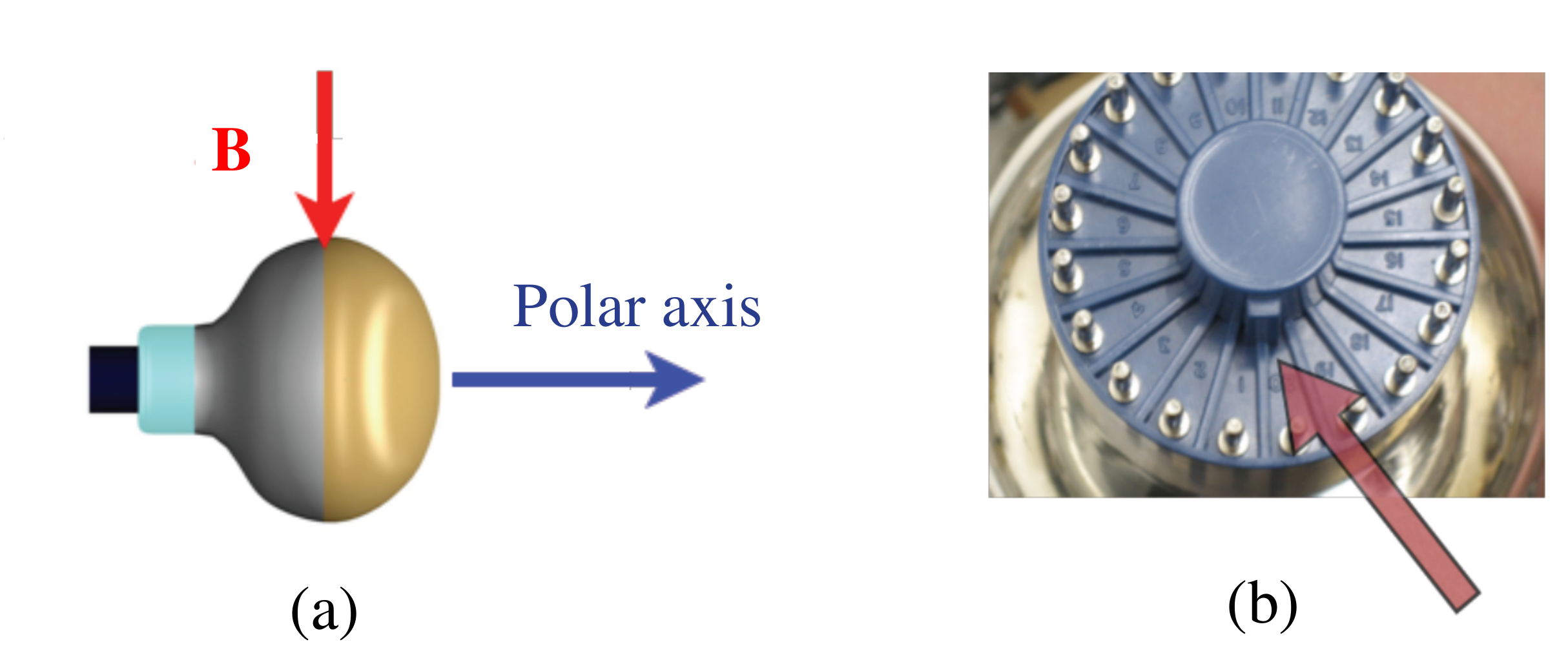}
\caption{(a) Orientation of the PMT with respect to the local magnetic field for 
evaluating response.  (b) PMT reference direction: the polarized key between pin 1 and pin 20 indicated by the arrow is used as the PMT reference direction.}
\label{fig:MagneticSetup}
\end{figure}

We determined the effect of the local magnetic field on the response of a PMT by positioning the PMT with its polar axis perpendicular to the field, as shown in Figure~\ref{fig:MagneticSetup}(a). The PMT was mounted on a holder that could be rotated about the polar axis from outside of the dark box containing the PMT.  Light from a pulsing blue LED illuminated the entire photocathode, and its intensity could be controlled as needed in the study.  The charge associated with the PMT signal was measured with an analog-to-digital converter (ADC) as a function of the angle between the local magnetic field and a reference on the PMT, denoted by $\phi$.  In this study, the polarized key on the socket of the PMT, shown in Figure~\ref{fig:MagneticSetup}(b), was chosen as the reference direction.  
When this reference direction was parallel (normal) to the local field, $\phi$   
was $0^{\circ}$ ($90^{\circ}$).

\begin{figure}[htpb]
\centering
\includegraphics[height=3.in, width=0.675\textwidth, keepaspectratio=false]{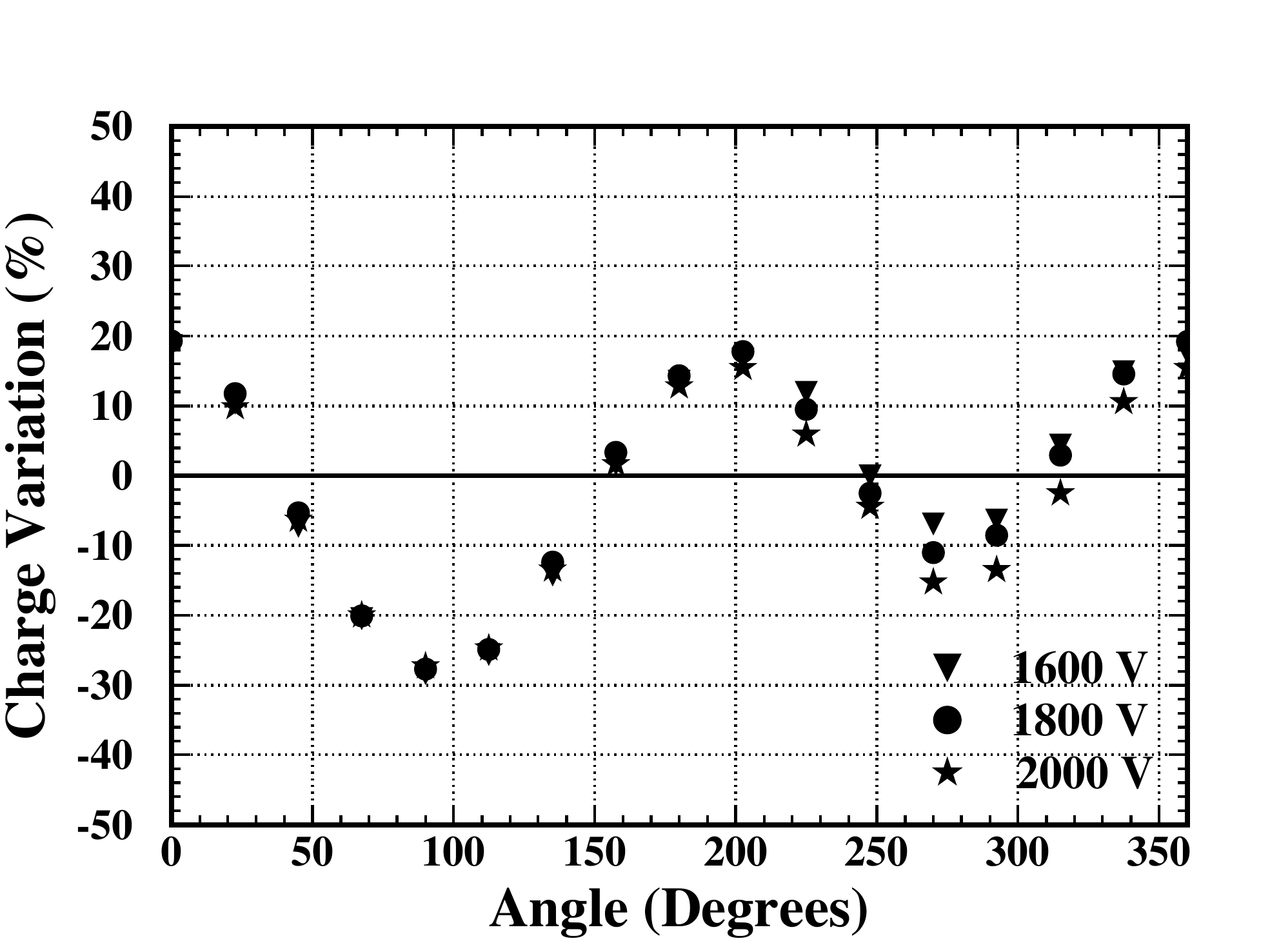}
\caption{Effect of magnetic field on collecting charge for an unshielded Electron Tube 9354KB PMT as a function of the angle between the PMT reference direction and the magnetic field.}
\label{fig:et5428_magnetic}
\end{figure}

\begin{figure}[htpb]
\centering
\includegraphics[height=3.in, width=0.675\textwidth, keepaspectratio=false]{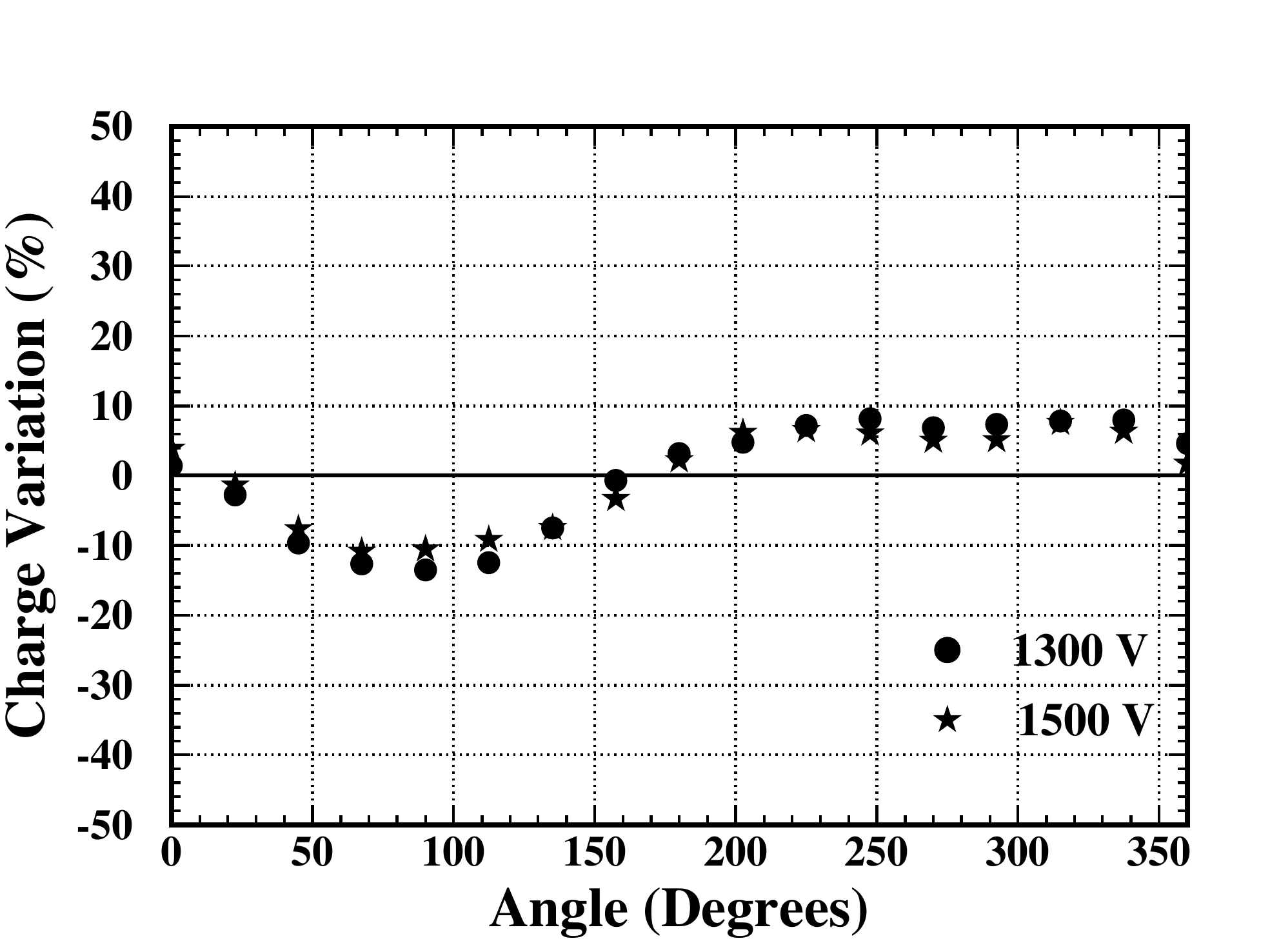}
\caption{Effect of magnetic field on collecting charge for an unshielded Hamamatsu R5912 PMT as a function of the angle between the PMT reference direction and the magnetic field.}
\label{fig:hama4327_magnetic}
\end{figure}

\begin{figure}[htpb]
\centering
\includegraphics[height=3.in, width=0.675\textwidth, keepaspectratio=false]{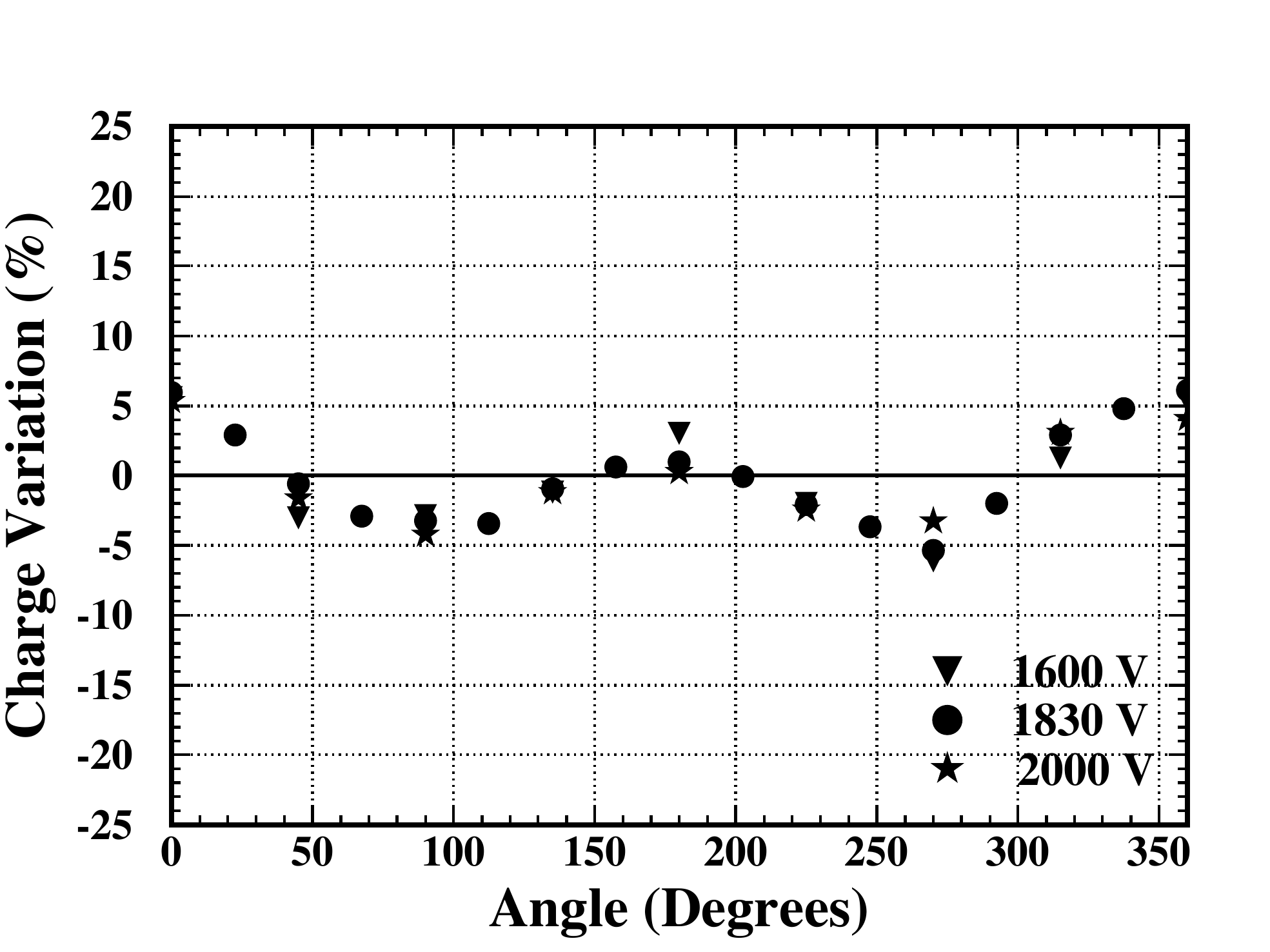}
\caption{Effect of magnetic field on collecting charge for an unshielded Photonis XP1806 PMT as a function of the angle between the PMT reference direction and the magnetic field.}
\label{fig:photonis1135_magnetic}
\end{figure}

The $\phi$-dependence of the charge variation at some operating high voltages 
for the Electron Tubes 9354KB, Hamamatsu R5912, and 
Photonis XP1806 PMTs is shown in Figures~\ref{fig:et5428_magnetic}, \ref{fig:hama4327_magnetic} and \ref{fig:photonis1135_magnetic} respectively.  
The charge variation, $\Delta Q$, is defined as 
\begin{equation}
\Delta Q(\phi) = \left(\frac{Q(\phi)-\overline{Q}}{\overline{Q}}\right) \times 100\%
\end{equation}
where $Q(\phi)$ is the amount of charge collected at angle $\phi$, and $\overline{Q}$ is the average amount of collected charge given by
\begin{equation}
\overline{Q} = \frac{1}{N}\sum_{i=1}^{N}Q(\phi_i)
\end{equation}
with $N$ being the number of measurements done at various $\phi$ values. 

\begin{figure}[htpb]
\centering
\includegraphics[height=1.7in, width=0.95\textwidth, keepaspectratio=false]{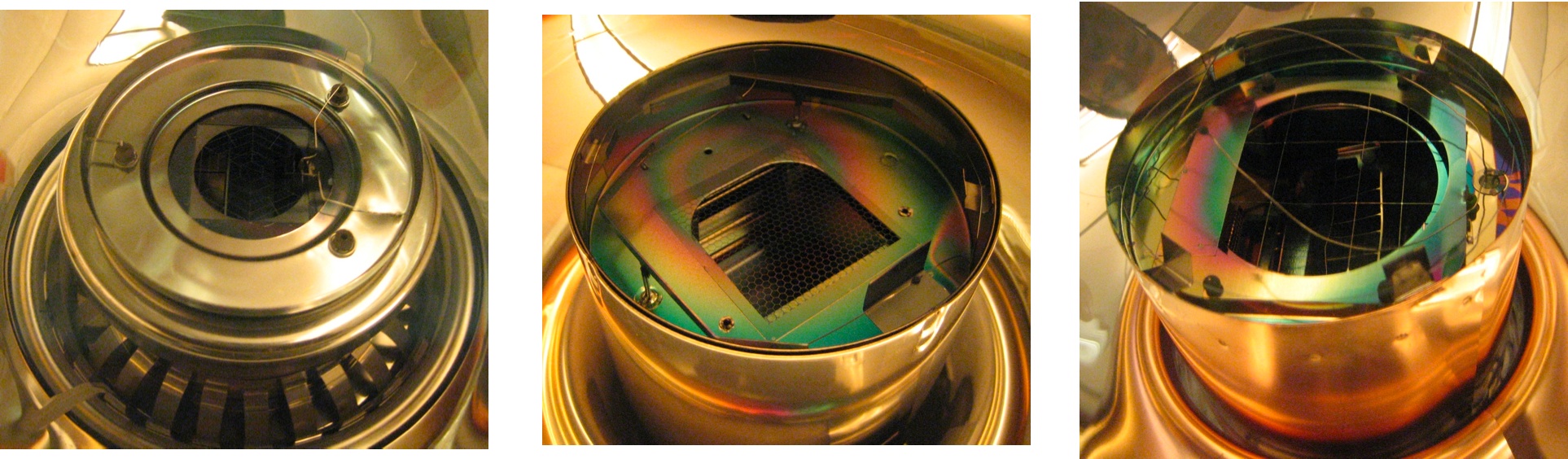}
\caption{Opening to the first dynode. From left to right: Electron Tubes 9354KB, Hamamatsu R5912, and Photonis XP1809.}
\label{fig:PMT_aperture}
\end{figure}

For each type of PMT the result is quite insensitive to the applied voltage.  The difference in the observed behaviour between the three PMT models is a consequence of the different designs of the dynode structure and focussing scheme.  Figure~\ref{fig:PMT_aperture} shows the opening to the first dynode for these three kinds of PMTs.    
Electron Tubes 9354KB has the smallest opening (most affected) whereas Photonis XP1806 has the largest opening (least affected).  

\subsection{Effect of Magnetic Field on Single-photoelectron (SPE) Spectrum}

\begin{figure}[htpb]
\centering
\includegraphics[height=2.in, width=0.45\textwidth, keepaspectratio=false]{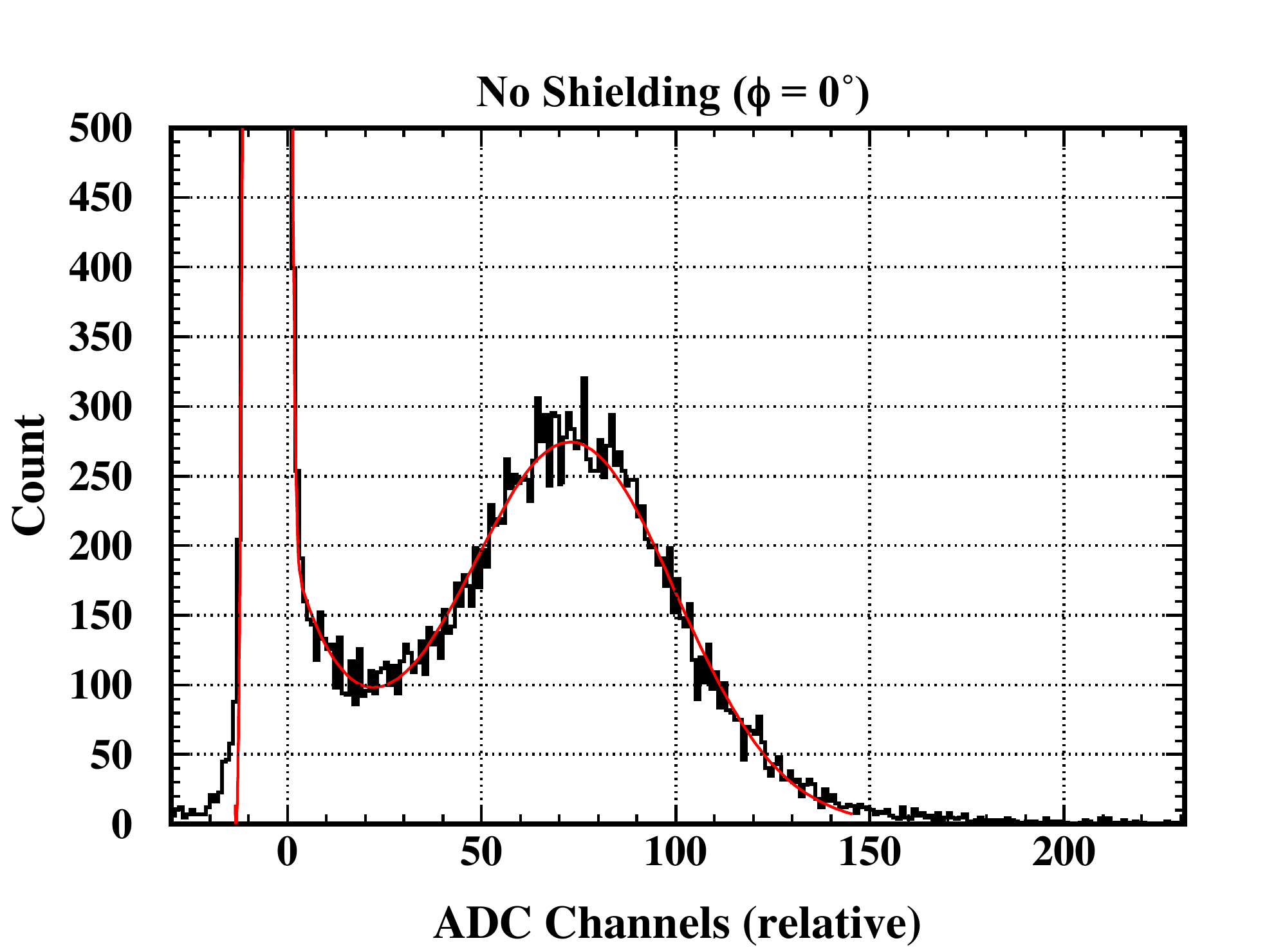}
\includegraphics[height=2.in, width=0.45\textwidth, keepaspectratio=false]{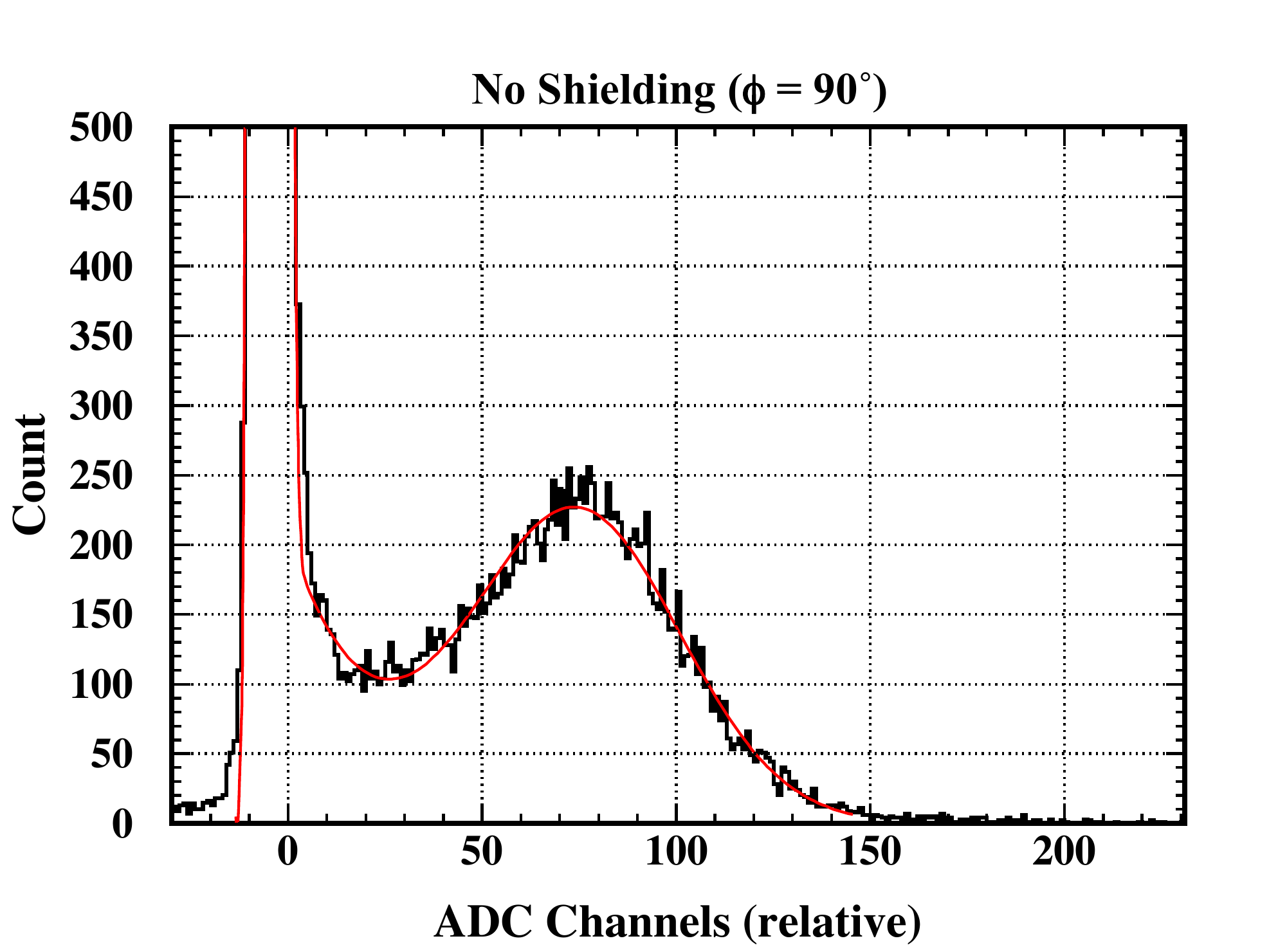}
\caption{Magnetic effect on the SPE distribution of a Hamamatsu R5912 without shielding. Figure on the left is obtained with the PMT reference direction parallel to the magnetic field ($\phi = 0^{\circ}$); figure on the right is determined with the magnetic field perpendicular to the PMT reference direction 
($\phi = 90^{\circ}$). The red curve in each figure is the best fit to the spectrum.}
\label{fig:hama4338_000_SPE}
\end{figure}

A revealing way to study the influence of the magnetic field on the detected charge is to use single-photoelectron events. 
By making this measurement without a magnetic shield we can directly determine how the shield affects both the gain and the collection efficiency of the PMT. 
The SPE spectrum can be obtained by reducing the light intensity on the photocathode until only about 10\% of the triggers produce a signal.  
From Poisson statistics about 95\% of the observed pulses are produced by SPEs.   
The spectrum of the charge collected at the anode under this condition is shown in Figure~\ref{fig:hama4338_000_SPE}.  Since the charge of a single electron is known {\it a priori}, the calibrated ADC channel of the SPE peak can be used to determine the absolute gain of the PMT.  Photoelectrons generated at the photocathode are directed to the first dynode by a set of focussing electrodes.  The configuration of the focussing elements and the effective area of the first dynode determines the collection efficiency.  
The Lorentz force, {\bf v} $\times$ {\bf B}, on the photoelectrons in the rather long distance between the photocathode and the first dynode causes some of them to be swept away from the first dynode and thereby reduces the collection efficiency. 
The number of events in the SPE distribution is proportional to the efficiency of collecting the produced SPE by the first dynode of the PMT. 


Using a Hamamatsu R5912 without a magnetic shield we have obtained SPE spectra with 
the PMT reference direction parallel ($\phi = 0^\circ$) and perpendicular 
($\phi = 90^\circ$) to the local magnetic field.
The observed ADC spectrum of each orientation, a convolution of the SPE distribution and the pedestal, is shown in Figure~\ref{fig:hama4338_000_SPE}. 
The observed spectrum was fitted to two Gaussian functions, one for the pedestal and the other one
for the SPE distribution, and an exponential function for modeling the non-Gaussian component of the
pedestal.
The gain , peak-to-valley ratio (P/V ratio) and charge resolution of the PMT were obtained from
the best fits which are displayed in Figure~\ref{fig:hama4338_000_SPE}.  
The results are tabulated in Table~\ref{tab:hama4338_000_SPE}.  
It is interesting to note that the change in gain between the two orientations is small. 
From Figure~\ref{fig:hama4327_magnetic}, we see a 15\% difference in the charge variation between these two angular positions; that is, the collection efficiency has changed by about 15\%.  This result suggests that the majority of the secondary electrons produced within the dynodes are not significantly affected by the magnetic field.  However, the peak-to-valley ratio gets worse when the magnetic effect is more pronounced.  
A series of measurements was made to determine how the collection efficiency in the presence of a magnetic field depends on the potential difference between the photocathode and first dynode. The results showed almost no effect on the collection efficiency when the potential difference was within 250 V of its nominal value, typically several hundred volts, for a gain of $10^7$.  


\vspace{0.5 cm}
\begin{table}[htdp]
\caption{SPE Spectra of a Hamamatsu R5912 in the presence of magnetic field.}
\begin{center}
\begin{tabular}{|c|c|c|c|}
\hline
PMT Orientation & Gain & P/V Ratio & Charge Resolution \\
\hline
\hline
$\varparallel$  magnetic field ($\phi = 0^{\circ}$)& $(1.030 \pm 0.003) \times 10^{7}$ & 2.77 $\pm$ 0.12 & 0.343 $\pm$ 0.003 \\
$\perp$ magnetic field ($\phi = 90^{\circ}$)& $(1.043 \pm 0.004) \times 10^{7}$ & 2.13 $\pm$ 0.11 & 0.333 $\pm$ 0.003 \\
\hline
\end{tabular}
\end{center}
\label{tab:hama4338_000_SPE}
\end{table}

\section{FINEMET magnetic shield}
In this section we describe the magnetic shield made of FINEMET, the material we used for reducing the effect of magnetic field on the Electron Tubes 9354KB, Hamamatsu R5912, and Photonis XP1806 PMTs.
 
\subsection{FINEMET${}^{\textregistered}$}
FINEMET is a trademark of Hitachi Materials Ltd. for an amorphous metallic alloy ribbon made up of nano-crystalline grains of iron, silicon, boron, and small amounts of copper and niobium \cite{Finemet}.
As shown in Table~\ref{tab:Finemet}, this novel soft magnetic alloy has high saturation 
magnetic flux density, low core loss, and high permeability over a wide range of frequencies. 
Its permeability varies by less than $\pm$10\% between $-50~^\circ$C and $150~^\circ$C, with very small aging effects.
We used 0.12-mm-thick FINEMET sheets for fabricating the magnetic shields.
This thin flexible material is made of a layer of 18 $\mu$m-thick 
FINEMET FT-3M tape laminated with 25-$\mu$m-thick hot-melt adhesive between two polyethylene terephthalate (PET) films. 
It can operate in a temperature range between $-40~^\circ$C and $80~^\circ$C.
It is interesting to note that one layer of FINEMET can reduce the local magnetic field of $4 \times 10^{-5}$ T by about 20 dB (from Figure 4 of reference ~\cite{Hitachi}).
  
\begin{table}[htdp]
\begin{center}
\begin{tabular}{|c|c|} \hline
Property							& Value \\ \hline
Density							& 7300 kg/m$^3$		\\
Resistivity							&  1.2 $\mu\Omega$-m	\\
Curie temperature					& ~570~$^\circ$C		\\   
Magnetic flux density (DC H = 800 A/m) 	& 1.23 T 				\\
Maximum relative permeability (DC)		& 70,000 				\\
\hline
\end{tabular}
\vspace{0.5cm}
\caption{Selected physical properties of Hitachi FINEMET FT-3M~\cite{Hitachi}.}
\label{tab:Finemet}
\end{center}
\end{table}

In addition to the physical properties provided by the manufacturer and listed in Table~\ref{tab:Finemet}, 
we have measured the amount of natural radioactivity of a few samples of FINEMET.  
Our results were $(18\pm5)$ ppb for the $^{238}$U series, $(15\pm5)$ ppb for the $^{232}$Th series, and 
$(4\pm2)$ ppm for K, which translate to decay rates of about 0.2 Bq/kg, 0.06 Bq/kg, 
and 0.1 Bq/kg for $^{238}$U, $^{232}$Th, and $^{40}$K respectively.
We have observed that if it is not stored in a ventilated location progenies of 
radon can accumulate gradually on the surfaces of the laminated FINEMET foil. 

\begin{figure}[htpb]
\centering
\includegraphics[height=3.in, keepaspectratio=true]{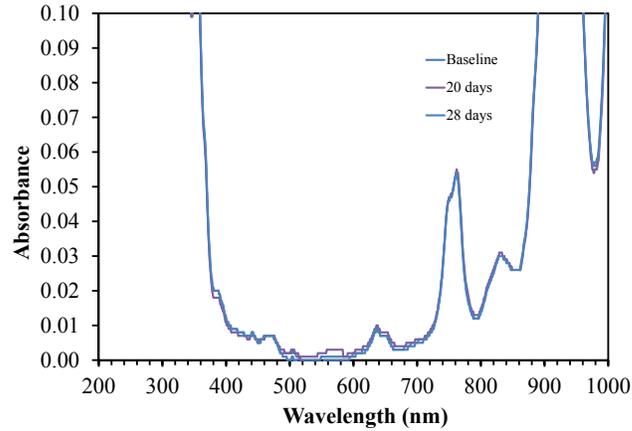}
\caption{Absorbance of mineral oil at $40~^\circ$C as a function of wavelength for a submerged FINEMET sample.}
\label{fig:Finemet-MO}
\end{figure}

\begin{figure}[htpb]
\centering
\includegraphics[height=3.in, keepaspectratio=true]{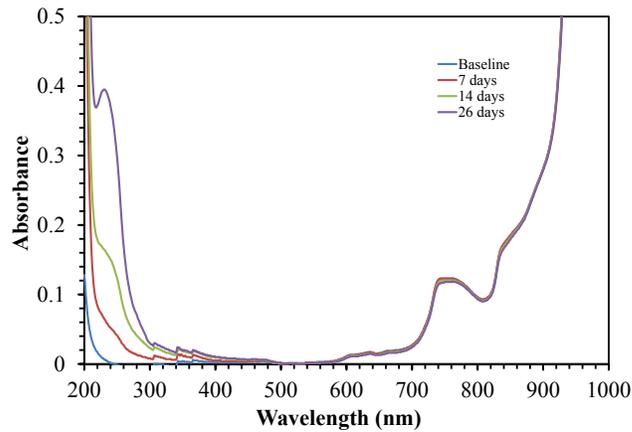}
\caption{Absorbance of purified water at $70~^\circ$C as a function of wavelength for a submerged FINEMET sample.}
\label{fig:Finemet-water}
\end{figure}

Besides radioactivity, we have also investigated the compatibility of FINEMET 
with mineral oil and 18-M$\Omega$ ultra-pure water, which are commonly used in nuclear and 
particle-physics experiments.
By submerging a piece of FINEMET sample with known dimensions in each liquid at 
elevated temperature for accelerated aging studies, the absorbance of the liquid was monitored as a function of time.  
For mineral oil at 40~$^\circ$C, with a ratio of contact surface to liquid volume (S/V) of 0.39 cm$^{-1}$, 
its absorbance as a function of wavelength and time is 
shown in Figure~\ref{fig:Finemet-MO}. 
The change in absorbance due to the presence of FINEMET is 
minute for wavelengths between 300 nm and 650 nm.  
The absorbance of purified water at 70~$^\circ$C for a FINEMET sample with a S/V of 0.32 cm$^{-1}$ is summarized in 
Figure~\ref{fig:Finemet-water}.
In the vicinity of 350 nm, the rate of change in absorbance levels off in about two weeks after a relatively fast degradation in the beginning. 
Note that the integrity of FINEMET has not changed in both liquids.

\subsection{Fabrication of Magnetic Shield}
\begin{figure}[htpb]
\centering
\includegraphics[height=2in, keepaspectratio=true]{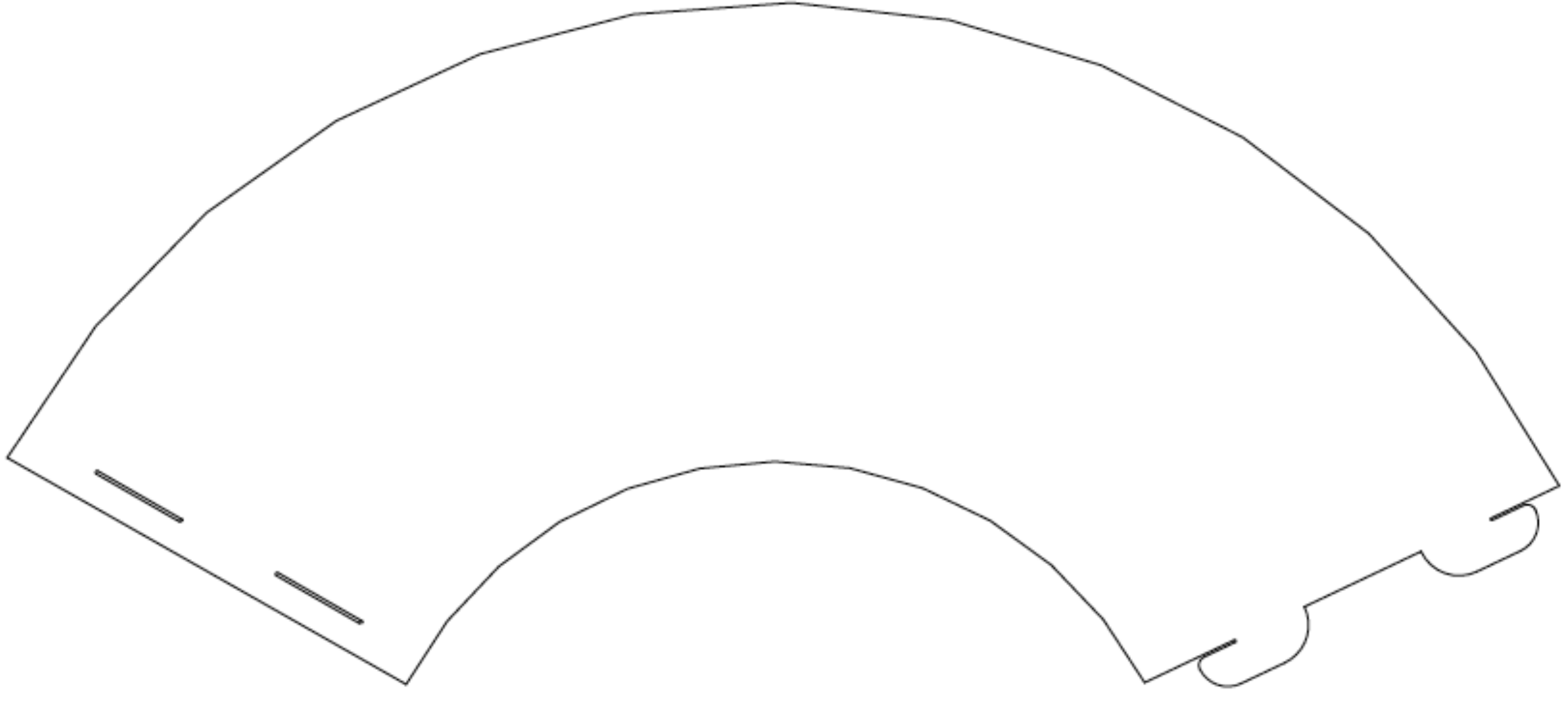}
\caption{Two-dimensional pattern of an unfolded truncated conical FINEMET magnetic shield.}
\label{fig:fan}
\end{figure}

\begin{figure}[htpb]
\centering
\includegraphics[height=2.5in, keepaspectratio=true]{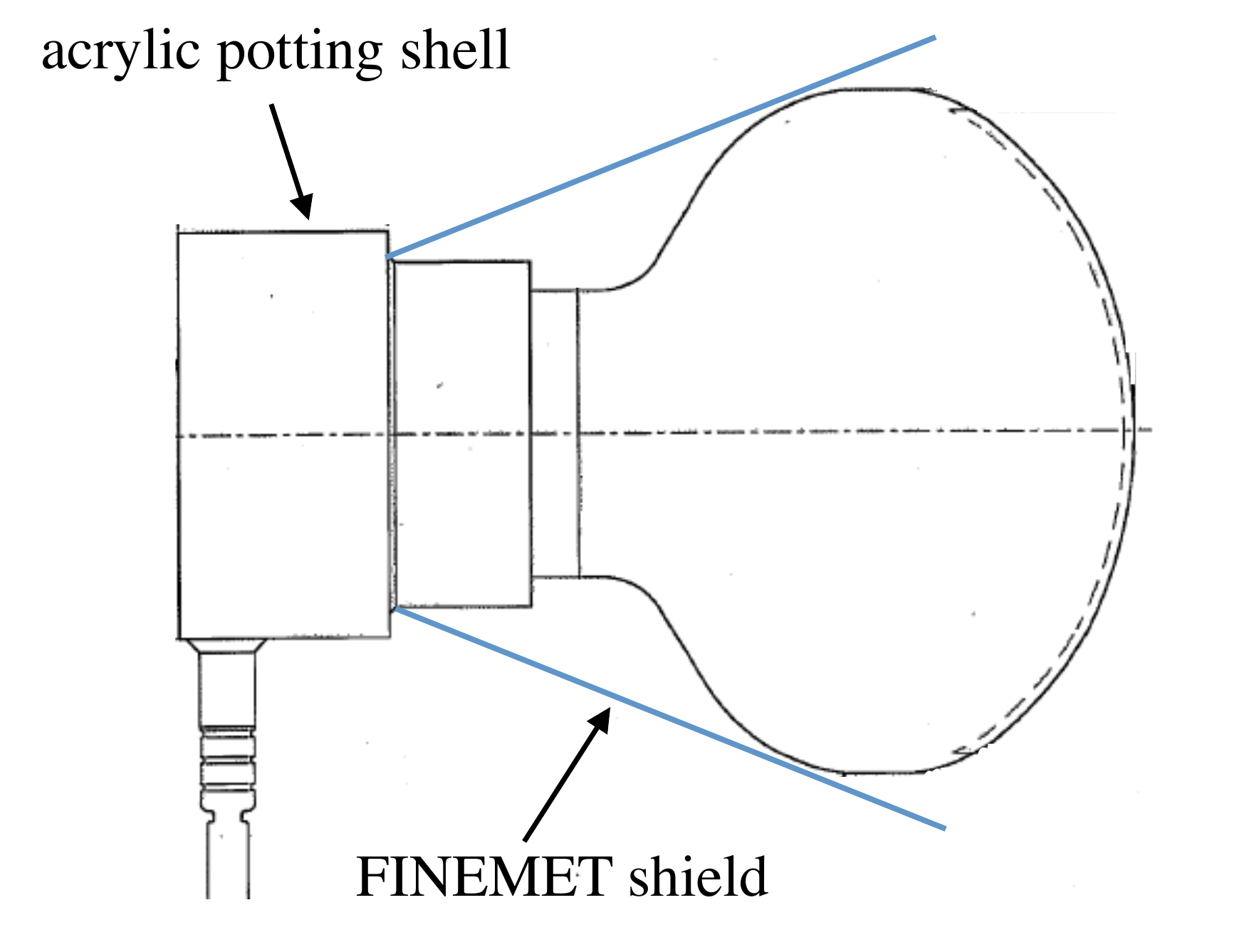}
\caption{Cross section of a Hamamatsu R5912 PMT used in Daya Bay with a truncated conical magnetic 
shield. The dashed line is the photocathode.}
\label{fig:ConeShield}
\end{figure}

As illustrated in Figure~\ref{fig:fan}, a truncated conical magnetic shield for the Hamamatsu R5912 PMTs used in the Daya Bay Experiment was realized by cutting a 
two-dimensional fan out of a piece of FINEMET foil with a water jet.  
The conically shaped shield was made by wrapping the fan around the PMT as shown in Figure~\ref{fig:ConeShield}.  It was firmly held in place between the equator of the glass bulb and the 
acrylic potting shell surrounding the base by tucking the tabs into the pre-cut slots. The shield is flexible, so any deformation of the cone can be taken out easily by simple manipulation of the surface.  
The weight of such a FINEMET shield for a typical 20-cm PMT is about 20 g.

\subsection{Performance of the FINEMET magnetic shield}
The effectiveness of the truncated conical FINEMET shield depends on the slant height of the cone. 
For the Hamamatsu R5912 assembly used in the Daya Bay experiment, the magnetic shield extends from the acrylic sealing shell to the equator of the bulb, where the photocathode ends.   We have evaluated the performance of the FINEMET magnetic shields with slant heights of 15.4 cm, 16.7 cm, 18.1 cm, 19.2 cm, and 23.0 cm, where the 23 cm-long shield extends to the plane tangent to the apex of the PMT bulb.  The nominal slant height of the shield used in Daya Bay is 15.4 cm, ending right at the contact point of the bulb near the equator. For the Photonis XP1806 PMT assembly we used shields with slant heights of 14.3 cm and 23.6 cm which extended to the equator and the apex of the PMT bulb respectively.  The corresponding slant heights are 14.5 cm and 22.5 cm for the Electron Tubes 5394KB PMT. 

\subsubsection{Improvement on the collection efficiency}

\begin{figure}[htpb]
\centering
\includegraphics[height=3.in, width=0.675\textwidth, keepaspectratio=false]{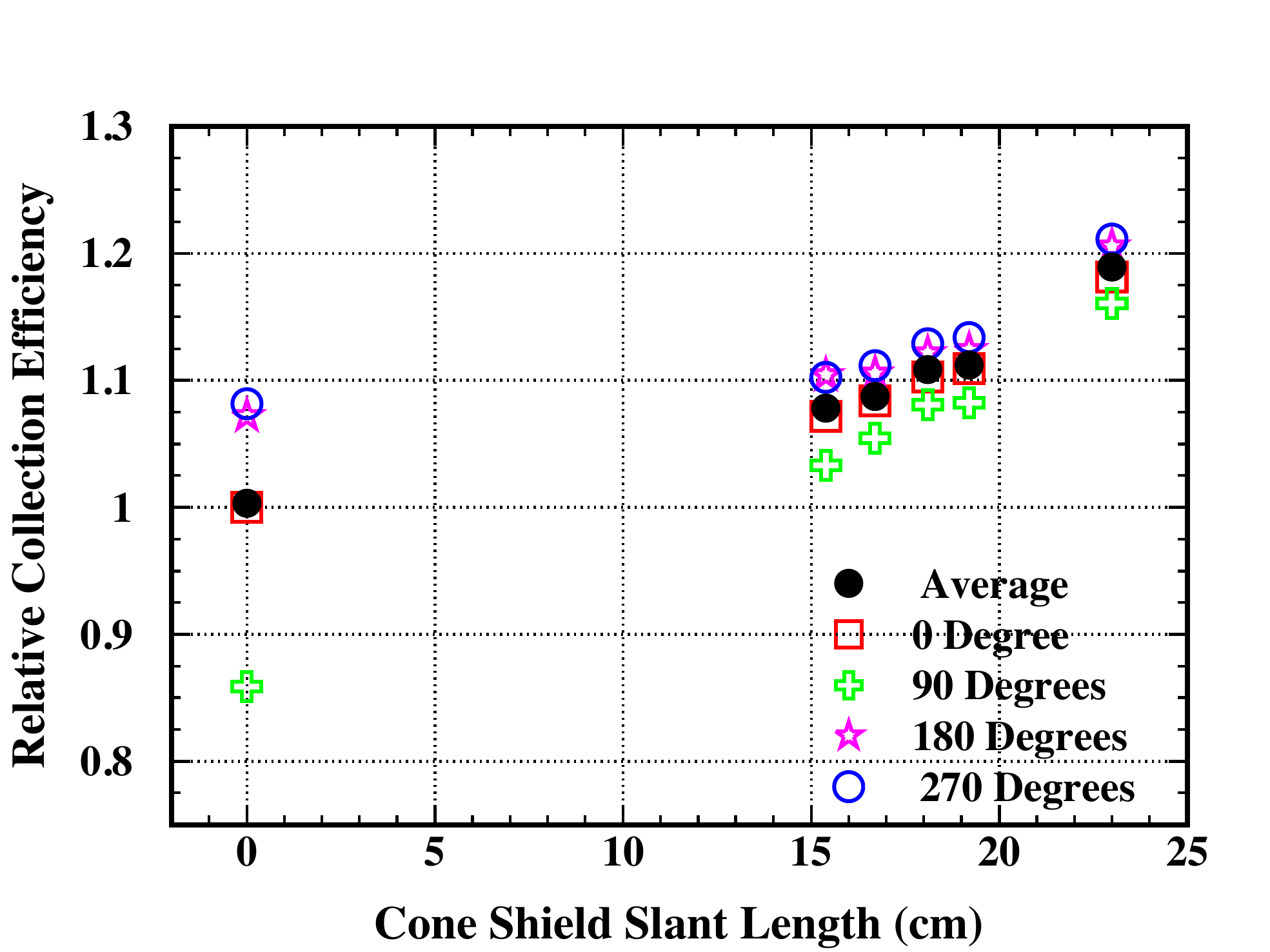}
\caption{Relative collection efficiency as a function of the slant height of truncated conical FINEMET shield for a Hamamatsu R5912
PMT oriented at different $\phi$ angles.}
\label{fig:hama4338_ceff}
\end{figure}

\begin{figure}[htpb]
\centering
\includegraphics[height=3.in, width=0.675\textwidth, keepaspectratio=false]{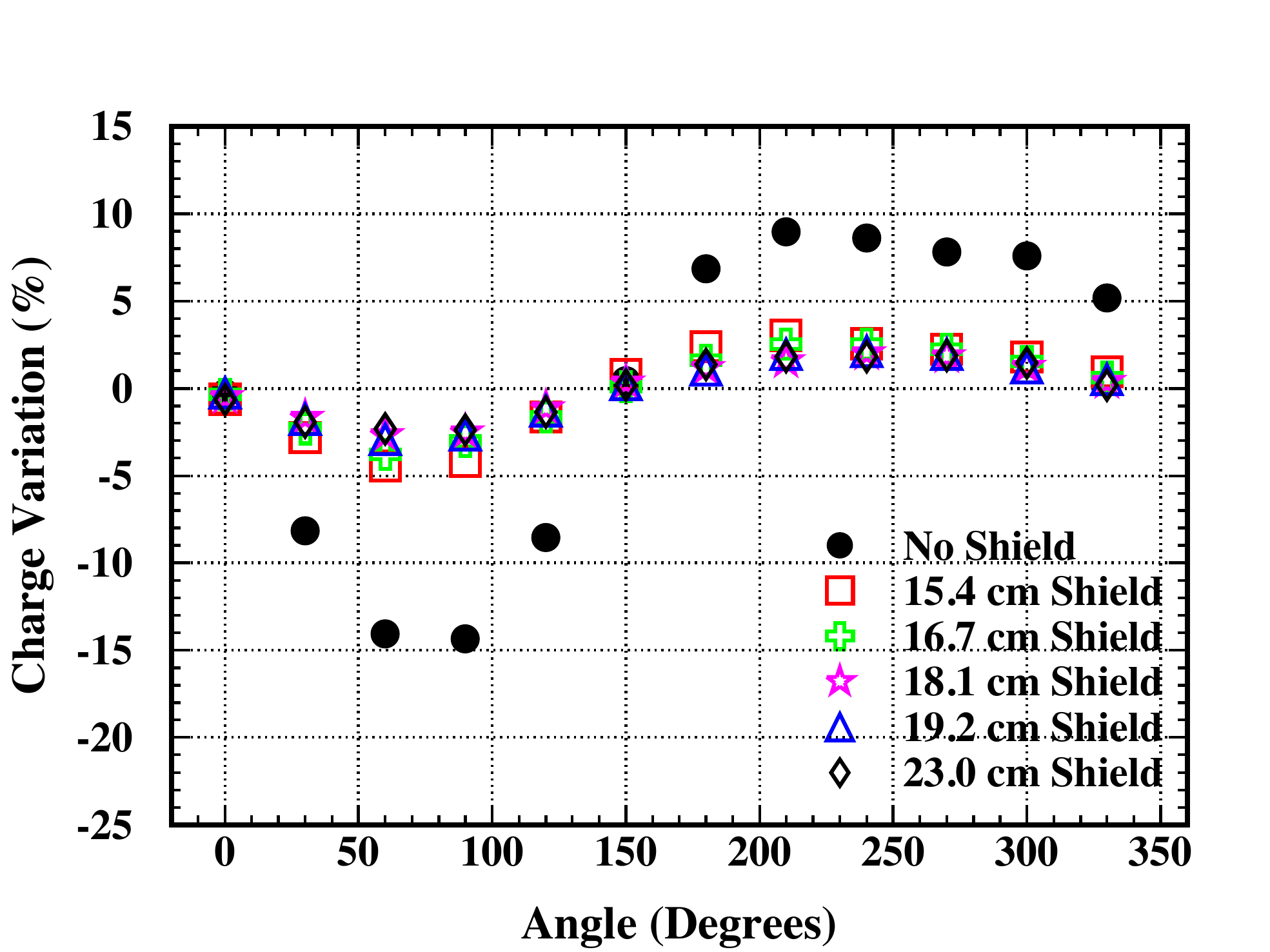}
\caption{Charge variation due to magnetic field for Hamamatsu R5912 PMT with FINEMET shields of different slant heights.}
\label{fig:hama4338_variation_all}
\end{figure}

With the polar axis of the PMT normal to the local magnetic field as shown in Figure~\ref{fig:MagneticSetup}, the amount of collected charge relative to the one without shielding at
$\phi = 0^{\circ}$  (relative collection efficiency) for the Hamamatsu R5912 PMT as a function of the slant height of the FINEMET shield for different $\phi$ angles is shown in Figure~\ref{fig:hama4338_ceff}. 
The magnetic shield clearly improves the collection of charge. 
Relative to the case of without any shielding, a FINEMET shield with a slant height of 15.4 cm (23.0 cm) improves
the average collection efficiency by 8\% (19\%). 
Furthermore, as shown in Figure~\ref{fig:hama4338_variation_all}, the charge variation as a function of the $\phi$ angle is reduced when the slant height increases. 
For example, without shielding, the peak-to-peak difference in the charge variation is about 23\%. 
A FINEMET shield with a slant height of 23.0 cm shrinks this difference to approximately 5\%.
Even a shield that covers only up to the equator (15.4 cm) allows us to achieve almost the same
performance.  This implies that the shield provides a less non-uniform collection efficiency to this 
level over the entire photocathode. 

\begin{figure}[htpb]
\centering
\includegraphics[height=3.in, width=0.675\textwidth, keepaspectratio=false]{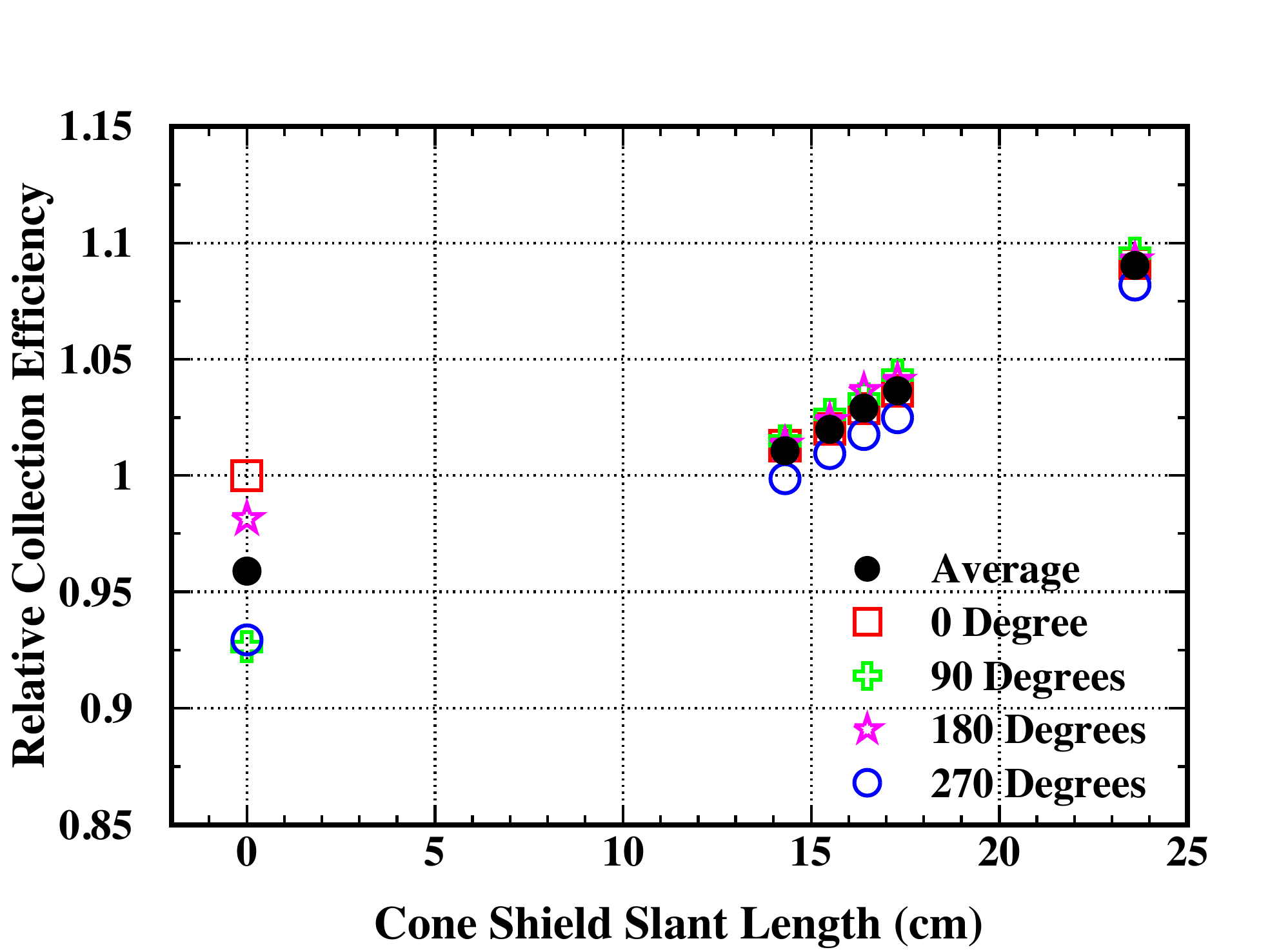}
\caption{Relative collection efficiency as a function of slant height of truncated conical FINEMET shield for a Photonis XP1806 PMT oriented at different $\phi$ angles.}
\label{fig:photonis1211_ceff}
\end{figure}
\begin{figure}[htpb]
\centering
\includegraphics[height=3.in, width=0.675\textwidth, keepaspectratio=false]{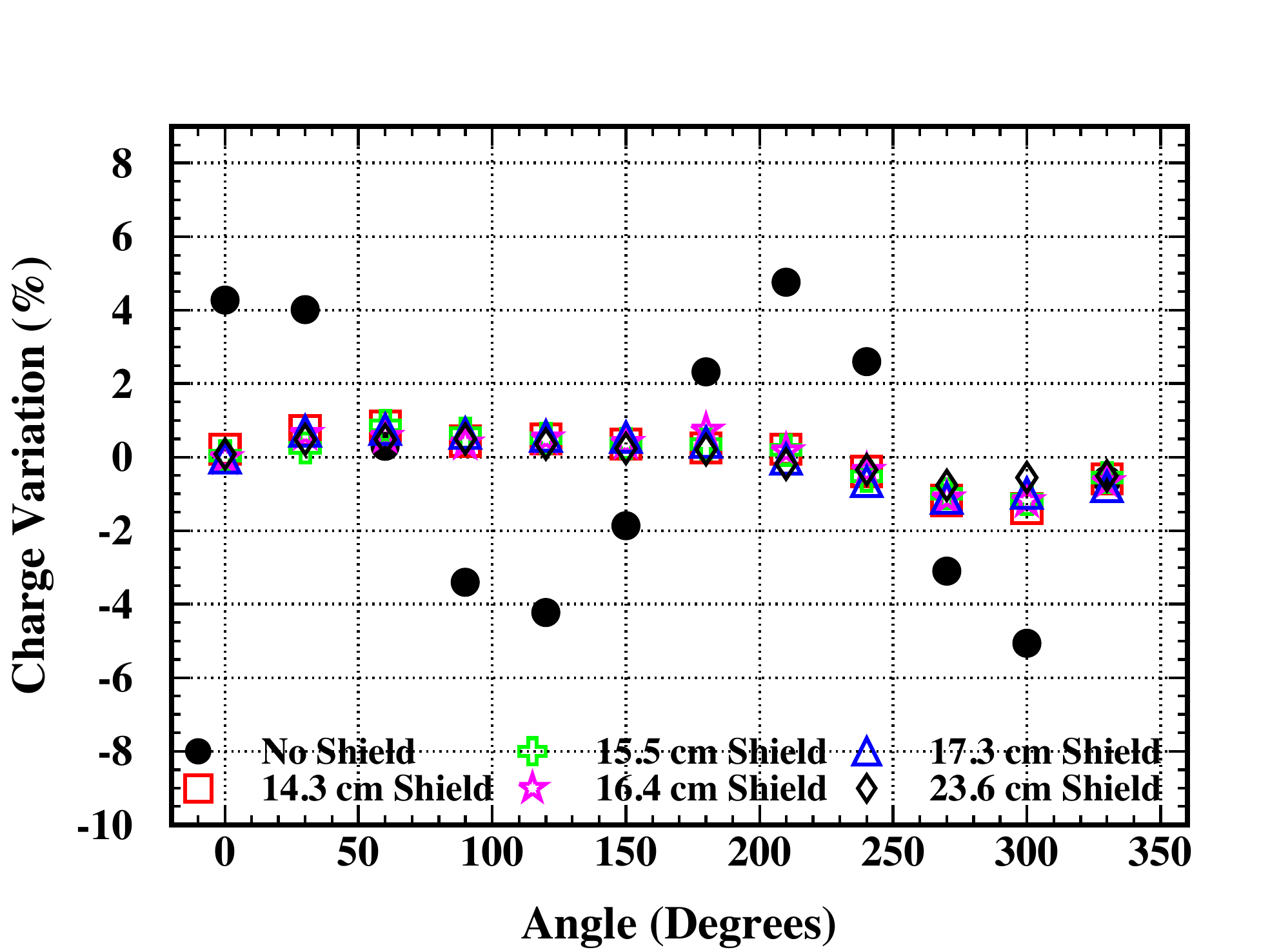}
\caption{Charge variation due to magnetic field for Photonis XP1806 PMT with 
FINEMET shields of different slant heights.}
\label{fig:photonis1211_variation_all}
\end{figure}
\begin{figure}[htpb]
\centering
\includegraphics[height=3.in, width=0.675\textwidth, keepaspectratio=false]{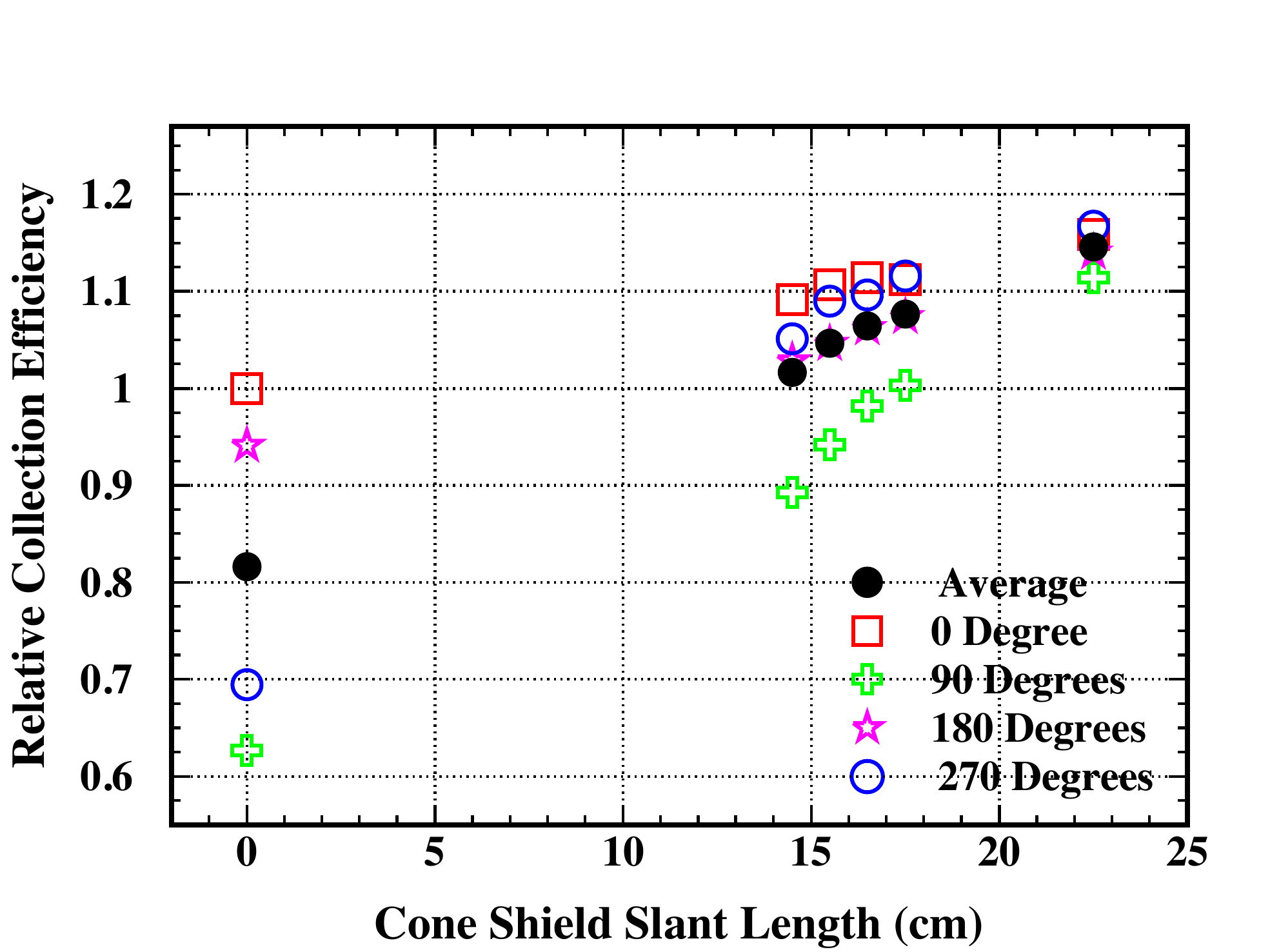}
\caption{Relative collection efficiency as a function of slant height of truncated conical FINEMET shield for an Electron Tubes 9354KB PMT oriented at different $\phi$ angles.}
\label{fig:et5428_ceff}
\end{figure}
\begin{figure}[htpb]
\centering
\includegraphics[height=3.in, width=0.675\textwidth, keepaspectratio=false]{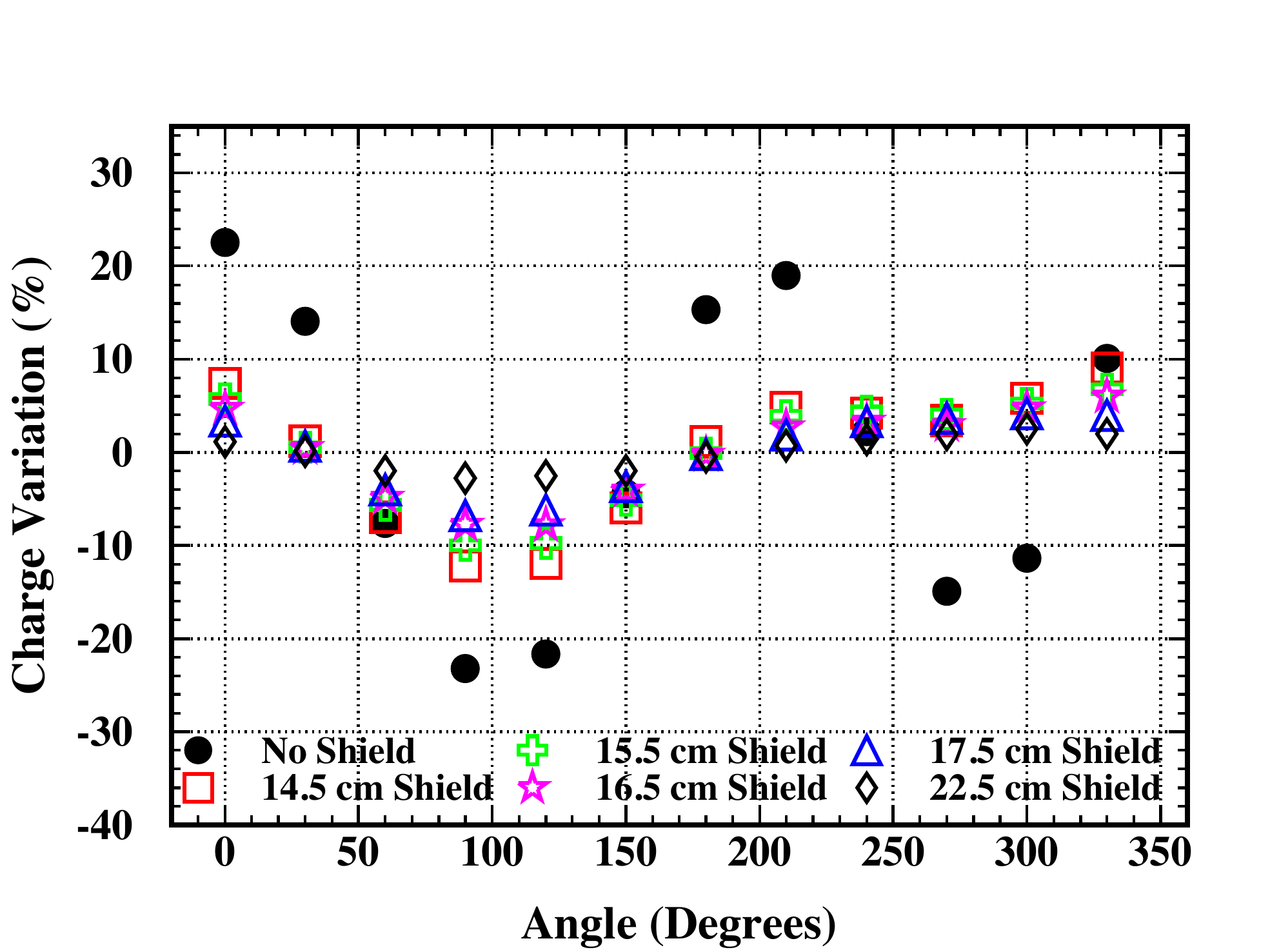}
\caption{Charge variation due to magnetic field for Electron Tubes 9354KB PMT with 
FINEMET shields of different slant heights.}
\label{fig:et5428_variation_all}
\end{figure}

Similarly, the results on collecting charge and minimizing the variation of the collection efficiency 
for the Photonis XP1806 and Electron Tubes 9354KB PMTs with FINEMET shields are depicted in 
Figures~\ref{fig:photonis1211_ceff}, \ref{fig:photonis1211_variation_all}, \ref{fig:et5428_ceff} 
and \ref{fig:et5428_variation_all}.
For the Photonis XP1806 PMT, on average, the collection efficiency is increased by 
about 5\% (14\%) for a shield with a slant height of 14.3 cm (23.6 cm), and 
the variation in the amount of charge collected is reduced from 10\% to 1.3\% which is close to our systematic limit. 
The most dramatic improvement is achieved with the Electron Tubes 9354KB PMT. 
In this case the average collection efficiency is up by about 25\% for a FINEMET shield 
extending to the equator and 40\% for shields reaching to the apex of the PMT bulb. 
The variation in the collection efficiency is reduced from 46\% with no shield to 
21\% (5\%) for a shield with a slant height of 14.5 cm (22.5 cm).

\subsubsection{Improvement on the SPE Spectrum}

In Table~\ref{tab:hama4338_shield_SPE} we summarize the results on the SPE spectrum for the 
Hamamatsu R5912 PMT with FINEMET shields of different sizes. 
Compared to the results for the case without shielding (Table~\ref{tab:hama4338_000_SPE}), 
we see that the shield can improve the gain by 5\%,  P/V ratio by up to 17\% (34\%)
for $\phi = 0^{\circ}$ ($90^{\circ}$), and charge resolution by less than 10\%. 
Furthermore, the variation of the P/V ratio due to different orientation of the PMT relative to the 
magnetic field is significantly reduced.
Consequently the systematic uncertainty in the charge measurement is minimized.

\begin{table}[htdp]
\caption{Characteristics of the SPE Spectrum for a Hamamatsu R5912 PMT with FINEMET magnetic shields of different slant heights.}
\begin{center}
\begin{tabular}{|c|c|c|c|}
\hline
PMT Reference Direction & Gain & P/V Ratio & Charge Resolution \\
\hline
\multicolumn{4}{|c|}{Slant height = 15.4 cm}\\
\hline
$\varparallel$ magnetic field ($\phi = 0^{\circ}$)& $(1.080 \pm 0.003) \times 10^{7}$ & 3.27 $\pm$ 0.18 & 0.316 $\pm$ 0.002 \\
$\perp$ magnetic field ($\phi = 90^{\circ}$)& $(1.092 \pm 0.003) \times 10^{7}$ & 2.86 $\pm$ 0.14 & 0.324 $\pm$ 0.003 \\
\hline
\hline
\multicolumn{4}{|c|}{Slant height = 18.1 cm}\\
\hline
$\varparallel$  magnetic field ($\phi = 0^{\circ}$)& $(1.091 \pm 0.003) \times 10^{7}$ & 3.17 $\pm$ 0.16 & 0.311 $\pm$ 0.002\\
$\perp$ magnetic field ($\phi = 90^{\circ}$)& $(1.088 \pm 0.03) \times 10^{7}$ & 3.03 $\pm$ 0.15 & 0.318 $\pm$ 0.003 \\
\hline
\hline
\multicolumn{4}{|c|}{Slant height = 23.0 cm}\\
\hline
$\varparallel$  magnetic field ($\phi = 0^{\circ}$)& $(1.081 \pm 0.003) \times 10^{7}$ & 3.20 $\pm$ 0.16 & 0.315 $\pm$ 0.002 \\
$\perp$ magnetic field ($\phi = 90^{\circ}$)& $(1.085 \pm 0.003) \times 10^{7}$ & 2.93 $\pm$ 0.14 & 0.311 $\pm$ 0.002 \\
\hline
\end{tabular}
\end{center}
\label{tab:hama4338_shield_SPE}
\end{table}%

\subsubsection{Position-dependence of PMT Response}
\begin{figure}[htpb]
\centering
\includegraphics[height=2.5in, width=1.\textwidth, keepaspectratio=true]{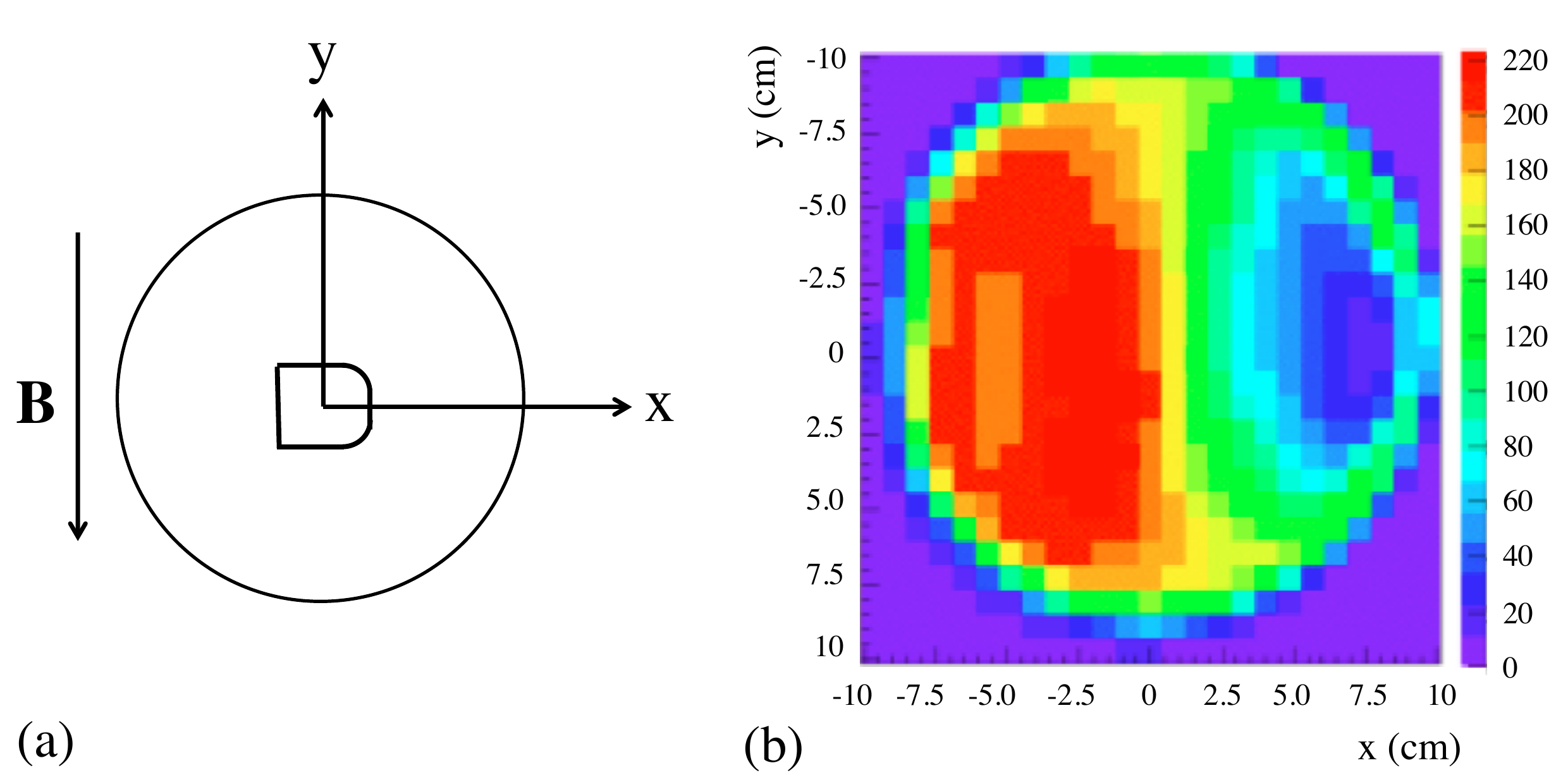}
\caption{(a) Orientation of the photocathode of a Hamamatsu R5912 PMT with respect to the local magnetic field. The x-y plane is normal to the polar axis of the PMT. The magnetic field is along the
negative-y direction.
(b) Position-dependent response of a Hamamatsu R5912 PMT with no magnetic shield.
The colored scale is the average number of detected PEs.}
\label{fig:uiuc1}
\end{figure}

We determined the response of a Hammamatsu R5912 PMT at each point of a grid on the photocathode by measuring the PE spectrum with a pulsing LED~\cite{UIUC}.
The orientation of the PMT is shown in Figure~\ref{fig:uiuc1}.
We split the light pulse from the LED into two parts with optical fibers. One part of the LED light was sent to a reference PMT to monitor the intensity of the LED, and the other part to the photocathode of the Hamamatsu PMT.  A collimator was used to limit the spot size of the incident light at the photocathode to a circle with a radius of 2 mm. 
The LED intensity was set to produce at least 20 photoelectrons per pulse on the photocathode.  Runs of ten seconds duration at a pulse rate of 100 Hz were used to measure the response of the PMT for every point on the photocathode.  An ADC was gated by the pulse generator used to drive the LED. A computer recorded the integrated counts in each ADC channel for each run.  The light source was moved under computer control to the next point on the grid until the whole surface of the photocathode was scanned.  

\begin{figure}[htpb]
\centering
\includegraphics[height=2.5in, width=1.\textwidth, keepaspectratio=true]{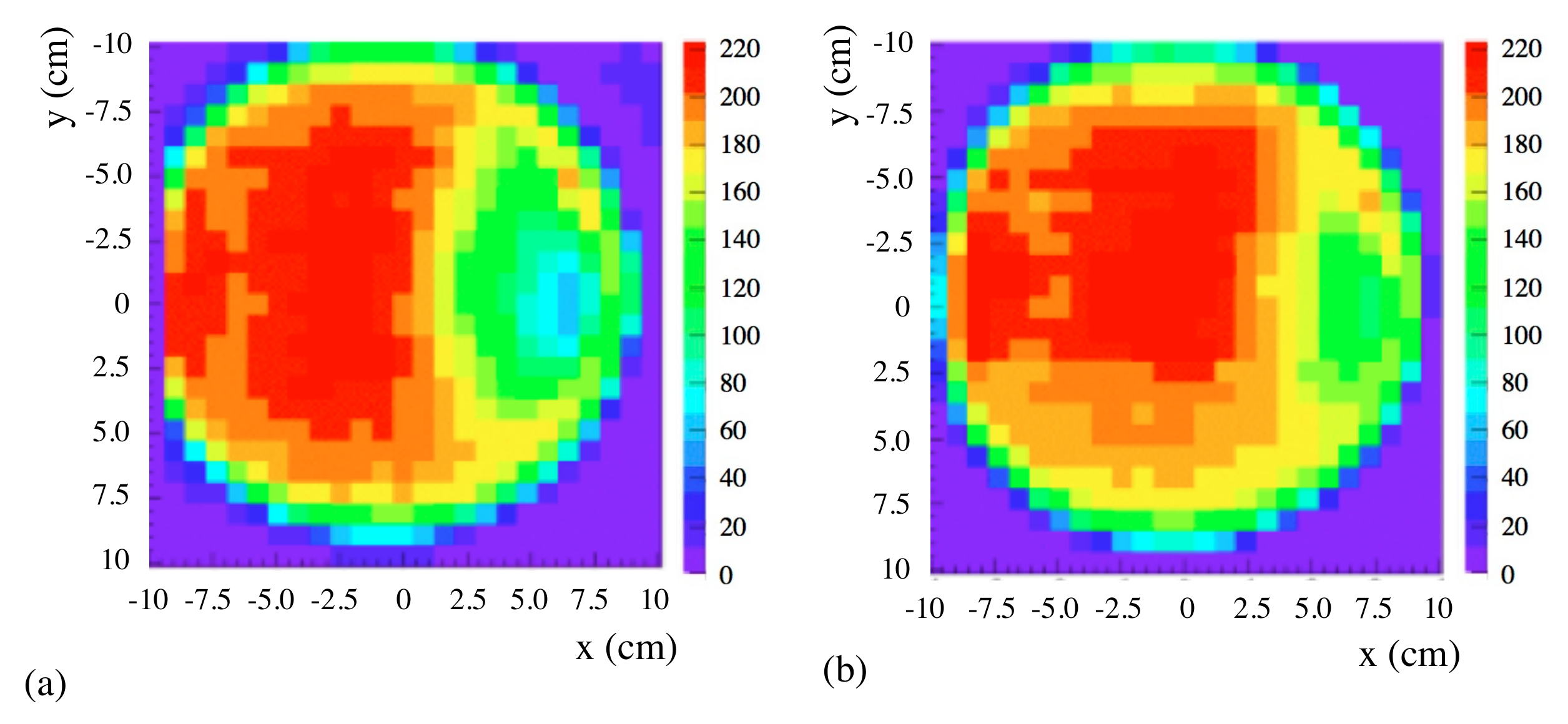}
\caption{Position-dependent response of a Hamamatsu R5912 PMT with a FINEMET shield
(a) with a slant height of 15.4 cm, (b) with a slant height of 20.5 cm. The colored scale is the average number of detected PEs.}
\label{fig:uiuc4}
\end{figure} 

In the absence of any magnetic field we expect rather uniform collection efficiency over the photocathode.  However, the Earth's magnetic field in the transverse direction creates a position-dependent response. In this case the field acting on a PMT with no shielding produces large non-uniformity as shown in Figure~\ref{fig:uiuc1}.  The lowest collection
efficiency occurs in a region centered at x $\approx$ 6 cm and y = 0 cm. In addition, the response of the PMT is almost symmetric about the x-axis.  This result is a direct consequence of the fact that the internal structure of the PMT is symmetric about the x-axis, so that with the magnetic field in the negative y-direction the response of the PMT should be symmetric across the x-axis.  This symmetry should not exist for the y-axis because the internal structure of the PMT is different across the y-axis, and also because a magnetic field in the negative y-direction will affect the motion of electrons in the x-direction.


A FINEMET shield around the PMT makes the response more uniform over the whole photocathode. The results for the shields with slant height of 15.4 cm and 20.5 cm are presented in Figure~\ref{fig:uiuc4}. 
The 20.5-cm shield has a more uniform response than the 15.4-cm shield, although even with this long shield there are still significant residual non-uniformities.
However, both of these shields clearly produce a better performance for the PMT than that obtained with no shield. 
  
\section{Conclusions}

We have presented the performance of a truncated conical magnetic shield made of FINEMET. Mechanically, it is thin, light, flexible, relative cheap, and easy to handle. In addition, 
the laminated FINEMET sheet is compatible with mineral oil and purified water. 
We have demonstrated that FINEMET is very effective in shielding large-aperture PMTs against magnetic field.
As a result, the collection efficiency of the PMT is improved and the response of the PMT
is more uniform across the entire photocathode.

\section{Acknowledgements}
We would like to express our gratitude to the technical staffs of the Lawrence 
Berkeley National Laboratory for their excellent support. 
This work was partially supported by the Director, Office of Science, Office of Basic Energy Sciences, Office of High Energy Physics, of the U.S. Department of Energy under Contract No. DE-AC02-05CH11231and DE-AC02-98CH10886, the U.S. National Science Foundation, the Research Grant Council of the Hong Kong Special Administrative Region, China (Project Nos. CUHK 1/07C and CUHK3/CRF/10).
J.N. and J. W. were also partially supported by the Department of Physics, Chinese University of Hong Kong.


\begin{thebibliography}{99}

\bibitem{pmt_hdbk}{Hamamatsu Photonics, "Photomultiplier Tubes: Basics and Applications", 3rd Edition, (2007).}

\bibitem{dayabay}{X. Guo {\it et. al.} (Daya Bay Collaboration), hep-ex/0701029 (2007).}

\bibitem{et9354}{Electron Tubes 9354KB Data Sheet, Electron Tubes Inc., Sept., 2002.}

\bibitem{hama5912}{Hamamatsu R5912 Data Sheet, Hamamatsu Co., Sept., 1998.}

\bibitem{xp1806}{Photonis Hemispherical PMT Data Sheet, Photonis Co, Sept., 2006.}

\bibitem{Finemet}
R. Lebourgeois {\it et al.}, J. Magn. Magn. Mater. {\bf 254-255}, 191 (2003).

\bibitem{Hitachi}
Catalog number HL-FM16-F, Hitachi Metals, Ltd. (http://www.hitachi-metals.co.jp).

\bibitem{UIUC}
T. Koblesky {\it et. al.}, Nucl. Instrum. Method A{\bf670}, 40 (2012).

\end{thebibliography}
\end{document}